\documentstyle[12pt,openbib]{article}
\begin{document}
\title{\bf The connection between the
nuclear matter mean-field equation of state and the quark  and gluon
condensates at high density.}
\vskip .3cm
\author{M. Malheiro $^{1}$ \thanks{Partially supported by CNPq of Brasil,
Present address: Department of Physics and Astronomy, University
of Maryland, College Park, Maryland 20742, USA}, Mira Dey $^{2}$
\thanks{Supported by CAPES of Brasil, work supported in part by
DST grant no. SP/S2/K04/93, Govt. of India, on leave from Lady
Brabourne College, Calcutta 700 017, India},  A. Delfino $^{1}$
\thanks{Partially supported by CNPq of Brasil},  and Jishnu Dey
$^{3}$ \thanks{Supported by FAPESP of Brasil, work supported in
part by DST grant no. SP/S2/K04/93, Govt.  of India, on leave
from W. Bengal Senior Education Service, India}\\
\\ $^{1}$ Instituto de F\'\i sica,
Universidade Federal Fluminense, \\ 24210-340, Niter\'oi,
R. J., Brasil
\\${^2}$ Instituto de F\'\i sica Te\'orica,
Universidade Estadual Paulista, \\ 01405-900, S\~ao Paulo,
S. P., Brasil \\
$^{3}$ Dept. de F\'isica, Instituto Tecnol\'ogico da Aeron\'autica,
\\ CTA, 12228-900, S\~ao Jos\'e Dos Campos, Brasil}
\vspace{.3 cm}
\date{\today }
\maketitle
{\it Abstract} :
	It is known now that chiral symmetry restoration requires the
meson-nucleon coupling to be density dependent in nuclear-matter mean-field
models. We further show that quite generally, the quark and gluon
condensates in medium are related to the trace of energy-momentum tensor
of nuclear matter and in these models  
the incompressibility, K, must be  less than thrice the chemical potential,
$\mu$. In the critical density $\rho _c$, the gluon condensate
is only reduced by $20\%$ indicating a larger effective nucleon
mass.

\newpage
Most of the models used in nuclear physics do not have explicit
chiral invariance, being based on effective Lagrangians. However
we have shown \cite{dddm} that some of these so-called QHD
models by virtue of the density-dependent coupling constants
produce ${\langle \bar q q\rangle}_\rho \rightarrow 0$ at high
$\rho$ and T.  These are the Zimanyi-Moskowski models \cite{zm}.

We inspect the density variation of the condensate in  hadron
models. These models yield different effective nucleon mass - its variation
with density is also model dependent. The pivotal question is the restoration
of chiral symmetry at high density : is there any correlation between the
nuclear matter equation of state (EOS) and the change of the condensates in the medium ?

In \cite{dddm} we have shown that in the linear Walecka model \cite{wa},
the ratio of the quark condensate of the nucleon in the medium, to that of
the vacuum,
\begin{equation}
R_{\bar q q} = \frac {{{\langle N| \bar q q|N \rangle}_\rho}}{{\langle 0|
\bar q q |0 \rangle}_0}
\label{eq:rat}
\end{equation}
goes down linearly upto 1.5 times the nuclear matter density but then tends
to {\it increase} (see Fig.1), contrary to expectations based on ideas of chiral symmetry
restoration. It could be argued that one should not push the Walecka model
beyond its range of validity and explore it at high density.  However, the
same kind of result is obtained by Li and Ko \cite{lk} very recently from
Bonn potential in the relativistic Dirac-Brueckner approach.  The problem is
not with the mean field approach which {\it should} work better at high
density.  Li and Ko also found the results unpalatable and conclude that
effects like dependence of meson - nucleon coupling constants on the current
quark mass may be crucial in obtaining a reliable result for the density
dependence of the quark condensate. Indeed we found \cite{dddm} contrary
result for variants of the Walecka model given by Zimanyi and Moszkowski (see
\cite{zm}). These models are labelled by ZM and ZM3  where the scalar  and
the scalar-vector mesons interact non-linearly and couplings are $\rho$ -
dependent.  In ZM and ZM3, the effective nucleon remains massive but
${\langle \bar q q\rangle}$ goes to zero at high $\rho$ (in Fig. 1 we
only show Walecka and ZM3 models. The results are
similar for ZM  \cite{dddm}). Thus we were forced
to conclude that the increase of $|{\langle \bar q q\rangle}|$ in the Walecka
model is not a fault
of the QHD approach or the mean field approximation but due to neglect of
non-linear coupling \cite{dddm}.

The present paper generalizes our previous work and  gives the following
important results : (1) We show that it is possible to relate the
condensate ratio, $R_{\bar q q}$ (eq.\ref{eq:rat}), to the trace of the
energy momentum tensor of nuclear matter, ${\cal E} - \, 3 P$, where $\cal E$
is the energy density and $P$ the pressure of the fluid. We remind the reader
that for a non -interacting fluid ${\cal E} > 3 P$ \cite{lan} but in presence
of non - electromagnetic interacion this is not necessarily true
; for example in the Walecka model the violation is due to
strong vector meson interaction.  The condition, ${\cal E} > 3
P$, however {\it must} be satisfied if $R_{\bar q q}$   is to
decrease, rather than increase. Of course we subscribe to the
common belief that {\it increase} of the condensate ratio with
density is unphysical. (2) In interacting nuclear models we must
not only have ${\cal E} > 3 P$ {\it but to ensure that the ratio
of the condensate $R_{ \bar q q}$ should fall monotonically we
must have a relation satisfied between their derivatives},
$\partial{\cal E} / \partial {\rho}\equiv \mu $ and  $\partial P
/ \partial {\rho} \equiv K/9$; namely $K < 3\mu$, K being the
nuclear matter incompressibility.  (3) The relation
that we find to relate $R_{ \bar q q}$ and ${\cal E} - \, 3 P$
is a quite general expression for mean-field relativistic nuclear
matter models. This is also valid for the non-linear Walecka
Model and still for chiral symmetric models as the chiral linear
$\sigma$-$\omega$ model.  (4) This relation, at the critical
density (when the quark condensate vanishes meaning a start of
the quark - gluon plasma phase), yields to the right bag
constant found from phenomenology \cite{ar}.  (5) We obtain
expressions for  gluon condensates in medium,
relating this to changes of the nucleon mass at high densities.

QCD, has a non-trivial vacuum with large non-perturbative condensates of
quarks and gluons. The non-zero value of the quark condensate is due to the
breaking of approximate chiral symmetry by the vacuum, otherwise enjoyed by
the QCD Hamiltonian, by virtue of the smallness of the quark mass $m_q$ :
\begin{equation}
H_{QCD} = H_0 + 2 m_q \bar q q
\label{eq:hqcd}
\end{equation}
the major part of the above being the chirally symmetric $H_0$. The
Hellmann-Feynman theorem relates the shift of the quark condensate from its
vacuum value to the nucleon sigma term and up and down quark
masses \cite{cfg2}:
\begin{equation}
2 m_q\int d^3x({\langle N|\bar q q|N\rangle}_N - {\langle 0|\bar q
q|0\rangle}_0)=2 m_q\int d^3x (\Delta q)_N= \sigma_N = m_q {\frac {d{M_N}}{d
m_q}}.
\label{eq:sigma1}
\end{equation}
where $|N\rangle $ is the free nucleon at rest and $(\Delta q)_N =  {\langle
N|\bar q q|N\rangle}_N - {\langle \bar q q\rangle}_0$. This term can be
looked upon as the "mass defect" of the nucleon due to $m_q \ne 0$ :
$\sigma_N = M_N-M_N^0$, $M_N^0 $ being the contribution from $H_0$ towards
$M_N$.

Using also the Hellmann-Feynman theorem and the relation of Gell-Mann, Oakes
and Renner, one can show \cite{dddm} that the condensate
ratio in medium to its vacuum value is
\begin{equation}
R_{\bar q q} = 1 - \rho \frac{\sigma_{eff}}{{m_\pi }^2{f_\pi
}^2}
\label{eq:fpimpi}
\end{equation}
with $f_\pi = 93$ MeV. And
\begin{equation}
\sigma _{eff} =\,\frac{ m_q}{\rho} {\frac {d{\cal E}}{d m_q}}
=\,\frac{\sigma_N}{\rho} {\frac {d{\cal E}}{d M_N}}
\label{eq:sigmaeff}
\end{equation}
is the effective $\sigma $-commutator for a nucleon in the nuclear medium.
The assumption is that the $m_q$ dependence of ${\cal E}$ comes through $M_N$.
If we neglect $\delta \cal E $, the nucleon kinetic and interaction energy
density in ${\cal E} = \rho M_N  + \delta {\cal E}$ we get $\sigma _{eff} =
\sigma_N$.

Using the expression for the energy density in the mean field approximation
(MFA) and calculating the derivatives in eq.(\ref{eq:sigmaeff}) , we
obtain a unified expression:

\begin{equation}
R_{\bar q q} = 1 - {\frac {\sigma_N}{{m_\pi }^2{f_\pi}^2}}
[\frac{{m_\sigma}^2} {{g^*_\sigma}^2} (M_N - M^*_N) + (1 + \alpha
)\frac{{m_\sigma}^2} {{g^*_\sigma}^2 M}(M_N - M^*_N)^2 - (1 +
\beta)\frac{{g_{\omega}^{\ast}}^{2}}{m^{2}_\omega M} \rho^2]  ,
\label{eq:qqrho}
\end{equation}
where the models in terms of $\alpha$ and $\beta$ are describes as :
Walecka\,($\alpha= \beta = 0$),
\,ZM\,($\alpha=\,1$\,and\,$\beta \,= \,0$)
and ZM3\,($\alpha = \beta\,= 1$\,) \cite{dddm}, \cite{zm}.

To our surprise we find that the expression between parenthesis in
eq.(\ref{eq:qqrho}) is equal to $({\cal E}-3P)/M_N$ which
(combining eqs. \ref{eq:fpimpi} and \ref{eq:sigmaeff}) can be written as

\begin{equation}
{\frac {d{\cal E}}{d M_N}}= \frac{({\cal E} -3 P)}{M_N} \, .
\label{eq:dEM}
\end{equation}

Therefore, a general expression for the quark
condensate in the medium arises:

\begin{equation}
\rho\,\sigma_{eff}\,=\,2 m_q({\langle N|\bar q q|N\rangle}_\rho -
{\langle 0| \bar q q|0\rangle}_0) =2 m_q\, \, (\Delta q)_\rho
=\frac {\sigma_N}{M_N}({\cal E} -3 P)
\label{eq:qqepp}
\end{equation}

or

\begin{equation}
R_{\bar q q}
= 1 - {\frac {\sigma_N}{{m_\pi }^2{f_\pi}^2}}
\frac{({\cal E} -3 P)}{M_N} \, .
\label{eq:ratio}
\end{equation}

Further, to connect the change of $R_{\bar q q}$ with $\rho$ we obtain

\begin{equation}
\frac {\partial R_{\bar q q}}{\partial \rho}
= \frac{1}{3} {\frac {\sigma_N}{{m_\pi }^2{f_\pi}^2}}
\frac{(K -3 \mu)}{M_N}
\label{eq:kmu}
\end{equation}

Thus the behaviour of the condensate is controlled not by the effective nucleon
 mass but by the EOS.  For a softer EOS there is
cancellation of the vector and the scalar fields. This makes $ {\cal E} > 3P$
and the condensate ratio decreases.  Monotonic decrease, however, depends on the
derivative of the EOS. From eq.(\ref{eq:kmu}) we find that  $K\,< 3 \mu
\equiv 3 {({\cal E} + P)}/{\rho}$ makes
the decrease monotonic. Otherwise the medium enhances the chiral breaking yielding
 models in which this happens to become unphysical. In other words, this means that
 the coupling constants {\it must } depend on $\rho$ to give a meaningful and soft
EOS valid at high $\rho$. Since $\mu $ is directly determined from $\rho$ the
condition $K\,< 3 \mu $ is the concrete criterion of softness for the EOS.
This is one of the central results of this paper.

Stiff EOS is not desirable in the context of phenomenology of
neutron stars. Also it leads to  very low effective nucleon
mass and high nuclear matter incompressibility.  We find the
absence of increasing the condensate ratio as another criterion
to rule out stiff EOS.

The models we used so far do not have chiral symmetry to start with and this
is not very satisfactory. To test whether chiral symmetric models may reveal
something different, we calculate the condensate in the chiral linear
$\sigma $- $\omega$ model \cite{Arim}. Again we find

\begin{equation}
{\frac {d{\cal E}}{d M_N}}= \frac{({\cal E} -3 P)}{M_N}
=[\frac{{m_\sigma}^2} {{g_\sigma}^2} \frac{(M_N^2 - M^{\ast{2}}_N)}{2
M_N}-\frac{{g_\omega}^{2}}{m^{2}_\omega M_N} \rho^2]\, \,.
\label{eq:chiral}
\end{equation}

The vector field , responsible for the $ \rho^2 $ term above,
controls the condensate at high $\rho$. 
Above the density,
 when $M^* \rightarrow 0$, the condensate ratio goes up ! It seems a common
feature that in any model, whether chirally symmetric or not, the vector
field (if not reduced by the density dependence of the coupling constant)
makes the EOS stiff and the condensate ratio increases.

Let us analyse the eq.(\ref{eq:qqepp}), connecting the trace of energy
momentum tensor, $T_\mu^\mu$,  via trace anomaly of QCD and a hydrodynamical
model of nuclear matter. Consider the latter first.
Embedding the nucleon in the infinite medium and treating the resulting
nuclear matter as perfect relativistic hydrodynamical fluid, the energy-momentum
tensor $T^{\mu \nu}$ (as we are dealing with a uniform and isolated system
with no heat flow-vector) is given by :

\begin{equation}
T^{\mu \nu} = ({\cal E} + P) u^{\mu} u^{\nu} - P g^{\mu \nu} \,.
\label{eq:Tuv}
\end{equation}
where ${\cal E}$ is the proper energy density, $P$ the hydrostatic
pressure and $u^{\mu}$ the four-velocity  of the fluid \cite{Fur} .

Write the fluid four-velocity vector $u^{\mu}$ in terms of 
the three-velocity as
\begin{equation}
u^{\mu} = \gamma (1,\vec v)  , \gamma = (1 - \vec v^2)^{-1/2}
\end{equation}
with $ u_{\mu} u^{\mu} = 1 $ . In the frame where nuclear matter is at
rest (comoving frame) $u^{\mu}=(1,0)$ the energy-momentum tensor is
diagonal 

\begin{equation}
{\langle T^{oo} \rangle}_{\rho} = {\cal E}\,\,,\,\,
{\langle T_{ij} \rangle} = P \, \delta_{ij}\, \, ,
\end{equation}
the expectation value of the trace of the energy-momentum tensor
$T_\mu ^\mu$ in nuclear matter from its vacuum value becomes
\begin{equation}
{\langle T_\mu^\mu \rangle}_{\rho} - {\langle T_\mu^\mu \rangle}_o
=  ({\cal E} -3 P)\, .
\label{eq:e3p}
\end{equation}

To connect with the microscopic picture we turn our attention to a recent
paper by Ji \cite {Xi}, the "QCD analysis of the Mass Structure of the
Nucleon". The energy-momentum
tensor is separated into a traceless $\bar T^{\mu \nu}$ and a trace $\hat{T}
^{\mu \nu}$ part contributing to $M_N$ as 
$(3/4) M_N$  and $(1/4) M_N$ respectively. The trace part is a well known
expression (in the leading order and  neglecting the anomalous dimension of
the mass term) :
\begin{equation}
\hat{T}^{\mu \nu}=\frac{1}{4}g^{\mu \nu} [ \bar \psi m\psi 
-\frac {9\alpha_s}{8\pi}{\bf G}^a_{\mu \nu}{\bf G}^{a \mu \nu}].
\label{eq:tmunu}
\end{equation}
where the second term is the anomaly term. Thus we have

\begin{equation}
\frac{M_N}{4}= {\langle N| H_a |N \rangle}+(\sigma_N + S)/4 \,,
\label{eq:M4}
\end{equation}
\\
where $H_a$ is the anomaly hamiltonian and $S$ is the strangeness content
of the nucleon, defined by
 $S=\int d^3x \,\, 
m_s (\langle N| \bar s s |N\rangle -\langle 0|
\bar s s |0\rangle ) \equiv \int d^3x \,\, 
m_s (\Delta s)_N$.

The divergence of the dilatation current, $\partial _{\mu}\,j^\mu _{dil}$, is
equal to the trace of $\hat {T}^{\mu
\nu}$ (eq. \ref{eq:tmunu}) which in the light sector becomes:

\begin{equation}
T_\mu^\mu = -\frac {9\alpha_s}{8\pi}{\bf G}^a_{\mu \nu}{\bf G}^{a \mu \nu}
+ m_u \bar u u+ m_d \bar d d + m_s \bar s s \, .
\label{eq:trace}
\end{equation}
It vanishes in the classical level for QCD in the chiral limit. Its
space part  vanishes for a
localised physical state such as $|N>$ :
\begin{equation}
\int d^3x \langle N|T^\mu _\mu| N\rangle = 
\int d^3x \langle
N|\partial _0 j^0_{dil} -\partial _i j^i_{dil}|N\rangle = \int d^3x \langle
N|\partial _0 j^0_{dil}|N\rangle 
\end{equation}
so that we get back $M_N$,
\begin{equation}
\int d^3x \langle N|T^\mu _\mu| N\rangle
 =\int d^3x \langle N|T_o^o|N\rangle = M_N.
\label{eq:tmn} 
\end{equation}

For a nucleon in the matter, this space part of  $\partial _\mu j^\mu_{dil}$ does
not vanish because of its infinite extent and we have the full trace on both
sides. Considering the translational invariance the volume of a single
nucleon in the medium is $1/\rho$ which appears on both sides and we have
\begin{equation}
({\cal E} -3 P)=(-9/8)(\Delta G)_\rho  + 2 m_q(\Delta q)_\rho
+ m_s(\Delta s)_\rho
\label{eq:ep}
\end{equation}
where $(\Delta G)_\rho \equiv {\langle N| \frac {\alpha_s}{\pi}{\bf G}^a_{\mu \nu}
{\bf G}^{a \mu \nu} |N \rangle}_\rho -  {\langle 0|\frac {\alpha_s}{\pi}
{\bf G}^a_{\mu \nu}{\bf G}^{a \mu \nu} |0 \rangle}$.
Similar to the fraction of the up and down quark mass term to $M_N$ ,

\begin{equation}
{\it b_1} = \sigma_N / M_N,
\label{eq:b1}
\end{equation}
if we define for the nucleon in the matter a fraction $b_2$,
\begin{equation}
{\it b_2} =\frac {\sigma_{eff}}{({\cal E} -3 P)/ \rho} = \frac{ 2 m_q(\Delta
q)_\rho}{({\cal E} -3 P)}\,\,.
\label{eq:b2}
\end{equation}
and impose the restriction,  $b_2 \equiv b_1$, so that
\begin{equation}
\rho\,\sigma_{eff}\,=\, 2 m_q(\Delta q)_\rho
= \frac{\sigma_N}{M_N}( {\cal E} -3 P)
\label{eq:e3p1}
\end{equation}
we arrive at the general expression given by (eq. \ref{eq:qqepp}).
The mean-field hadronic model calculations imply above relation
factoring out the medium dependence only in the trace energy of
the medium.

 Using this trace anomaly , one can relate the shift of the gluon condensate
in medium (from its vacuum value)\cite {cfg} ,\cite{cfg2}.
Combining the eqs.(\ref{eq:M4} and \ref{eq:tmn}) we obtain for the free
nucleon at rest:

\begin{equation}
M_N =\int d^3x (-\frac{9}{8})(\Delta G)_N + \sigma_N + S
=4 {\langle N| H_a |N \rangle} + \sigma_N+ S
\label{eq:ano}
\end{equation}
where $(\Delta G)_N $ is the shift of the gluon condensate in free nucleon
state relative to vacuum.

Analogously, defining a "gluon sigma" term $G_N$ given by
\begin{equation}
 G_N = \int d^3x (\Delta G)_N =\,-\frac{8}{9}(M_N - \sigma_N - S)\,
 =-\frac{32}{9} {\langle N| H_a |N \rangle}
\label{eq:gn}
\end{equation}
we find that in the two limits of $m_s$ (150 MeV and larger ),
studied in \cite{Xi}, where $S+\sigma_N=110$ or $160$ Mev
respectively, the corresponding values for $G_N$ are
$-737$ or $-693$ MeV. For $S=0$, the $G_N = -795$ MeV.

Following the same factorization as that of the light quarks we can write
the shift in strange quark condensate in the medium as
\begin{equation}
\rho\,S_{eff}\,=\,  m_s(\Delta s)_\rho
= \frac{S}{M_N}( {\cal E} -3 P)
\label{eq:qqs}
\end{equation}
where the density dependence comes again by the trace energy and S
is the strangeness content of the nucleon, the precise value of which is a
subject of some controversy \cite{Xi}. We obtain only a range for this change
depending on the estimation of this quantity, and in our calculations we
consider $S=0$ for which the shift in the gluon condensate is maximum.

The expressions for the quark condensates and 
the eqs.( \ref{eq:ep} and \ref{eq:gn}
) can be used to extract the "effective gluon sigma term" , $G_{eff}$,
for the nucleon embedded in the medium :
\begin{equation}
\rho \, G_{eff}\,=\,({\Delta G})_\rho 
= \frac{G_N}{M_N}({\cal
E}-3P)
\label{eq:gneff}
\end{equation}

The eq.(\ref{eq:ratio}) shows that at a
certain critical density $\rho_c$, it may happen that $R_{ \bar
q q}\rightarrow 0$. This corresponds ${\cal E} - \, 3 P \,
=\,4\,(171 MeV)^4 $. It is to be noted that even at this density the gluon
condensate in the nucleon is substantial :
\begin{equation}
{\langle N| \frac {\alpha_s}{\pi}{\bf G}^a_{\mu \nu}
{\bf G}^{a \mu \nu} |N \rangle}_\rho = (350)^4-3.4\,(171)^4 \sim (332 MeV)^4\,,
\end{equation}
which is only a reduction of $20\%$ , as we can see from fig.2. In 
normal density this reduction is even smaller and about $5\%$ using
the vacuum value 
$ {\langle 0|\frac {\alpha_s}{\pi}
{\bf G}^a_{\mu \nu}{\bf G}^{a \mu \nu} |0 \rangle}=(350 MeV)^4$.

Gluon contribution to the nucleon mass is more than the quark contribution
(510 MeV, as against 430 MeV, \cite{Xi}) and gluons seem to persist even at high
density. Quark part of nucleon mass in the medium might change but the
gluonic part remains. This explains why even at high densities
the nucleon mass can not be small and ZM3 model "effectively" reproduces this 
(this result is  reproduced in the non-linear Walecka model
 which also gives a good value  for the nuclear matter incompressibility).

Although the derivation of the eq.(\ref{eq:e3p1}) is very
general- uses only the trace of
the energy momentum tensor and the mean field approach which
 should work better at high density, when applied to
models we find that the condensate can behave very differently.
However, the following comments can be made with regard to some
model independent results:

(1) At the nuclear saturation density $\rho_0$, in any hadron model, we shall
have the same value of the condensate. The reason is obvious. At this $\rho_0$
the $P = 0$ and $ {\cal E} $ is made to be the same. Actually, with the
binding energy of -15.75 MeV and $M_N = 939$ MeV the condensate ratios
become, for example, $R_{\bar q q}=0.69$ and $R_{\bar g g}=0.95$ for $\sigma_N=45$
 Mev and S=0.
\\
(2) There exists a critical density $\rho _c$ for which the condensate goes
to zero. At this density $ {\cal E} - 3P = {m_\pi }^2{f_\pi}^2 M_N
/ \sigma_N $. This will imply a bag
constant $B^{1/4} = 171 MeV$. Notice this is the optimum bag constant if one
fits all the known hadrons in the manner of Aerts and Rafelski \cite{ar}.
After $\rho _c$ it is not realistic to picture nucleons as point particles.
The quark stucture is almost inevitable at this density.
\\
(3)  If there is scaling for the condensate
with the effective nucleon mass, $M_N^{\ast}$  vanishes when the
condensate vanishes. This is awkward for nuclear physicists since the meaning
of nuclei or nuclear matter composed of zero mass particles is ill-defined.
It is known that for a relativistic gas composed of non interacting particles
with rest mass $M_N$, ${{\cal E} - 3P}= \rho M_N (1-\frac{v^2}{c^2})^{1/2} $
 where v is the mean velocity of the particles \cite{lan}. This
means that for small densities   ${{\cal E} -
3P} \sim \rho M_N $ and we obtain the usual relations for the quark and
gluon condensates in the leading order approximation \cite{cfg} and \cite
{cfg2}. However, when the density and pressure increase
 $({\cal E} - 3P)< \rho  M_N $. If we define an
  effective mass
$M_N^* = M_N (1-\frac{v^2}{c^2})^{1/2}$, $M_N^*$ becomes very light when
${\cal E} \sim 3P $. But this leads to $R_{\bar q q} \rightarrow 1$ and
$R_{\bar g g}\rightarrow 1$ i.e. the condensates resume their vacuum values
which is opposite to our notion of chiral symmetry restoration. Thus we
stress that light effective nucleon mass,  {\it rather than restoring the
chiral symmetry, will enhance the symmetry breaking}.
\\
(4) Putting all the eqs. (\ref{eq:e3p1}, \ref{eq:qqs} and \ref{eq:gneff}), we
can write the effective quantities for the nucleon in the medium as

\begin{equation}
\rho \, \xi_{eff}
= \frac{\xi_N}{M_N}( {\cal E} -3 P)
\end{equation}
where $\xi_{eff}$ stands for the $\sigma_{eff}$, $G_{eff}$ or
$S_{eff}$. As long as the $({\cal E} -3P)/M_N < \rho$ and
{\it positive}, these effective quantities will decrease in the
medium. In Fig.3 we show that they are not far from the leading
order in the ZM3 model but in the Walecka model they become
unphysical. This is another illustration of the breakdown of the
Walecka model, where the $M^*_N$ is very light and the
condensates increase at high density.
\\
(5) The condition $K\,< 3 \mu $ allows us to conclude that the pressure
 can not increase fast with the density ( small incompressibility K at
  high $\rho$ ), which manifests that a soft EOS at high density is needed
  to expect that the medium will enhance chiral symmetry and the restoration
  occurs ( the case of ZM3 model ). This conclusion is also satisfied
 if we include the temperature effects, where now 
 $\mu \equiv \partial \Omega / \partial {\rho} $ where $\Omega$
stands for the grand-canonical potential.

It is important to point out that
in  this density analysis of the condensates our hadronic models
do not have pions in 2nd order.  In fact the pionic corrections will be there
- but as commented and observed in the reference \cite{Birse} -
a calculation with linear sigma model - their net effect should
be small.


\noindent

\newpage

{\bf Figure Captions}
\\
\\
Fig. 1: Ratio of the up-and-down quark condensate to the vacuum value and
the effective nucleon mass ratio $M^*$/M as a function
 of density for Walecka and ZM3
models.\\
Fig. 2: Ratio of the gluon condensate to the vacuum value for the models
 with S=0.\\
Fig. 3: The trace energy of nuclear matter divided by the nucleon
rest mass as a function of the density is plotted and compared  with
 the leading order approximation given by  $\rho / \rho_o$ (full line).

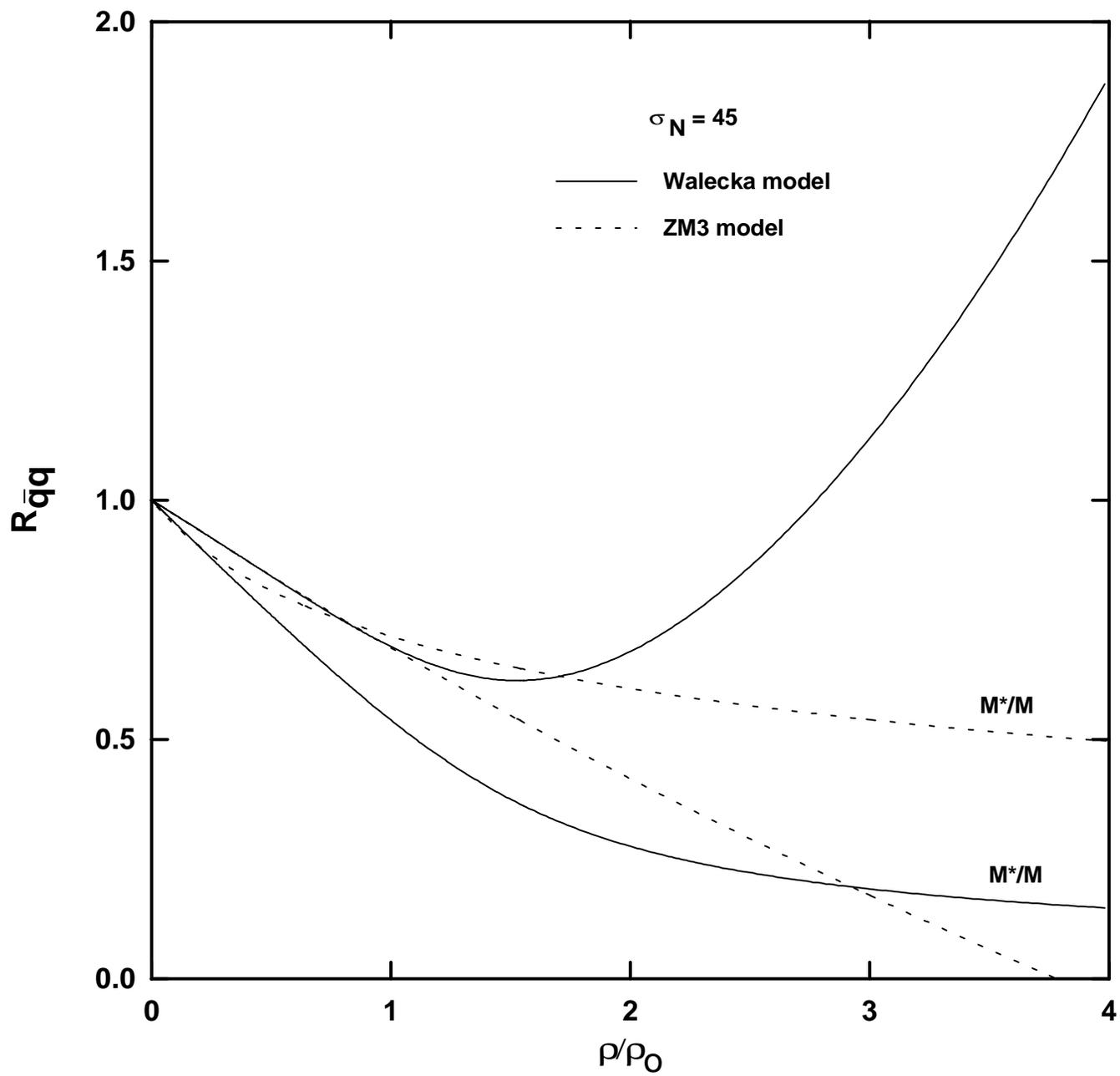
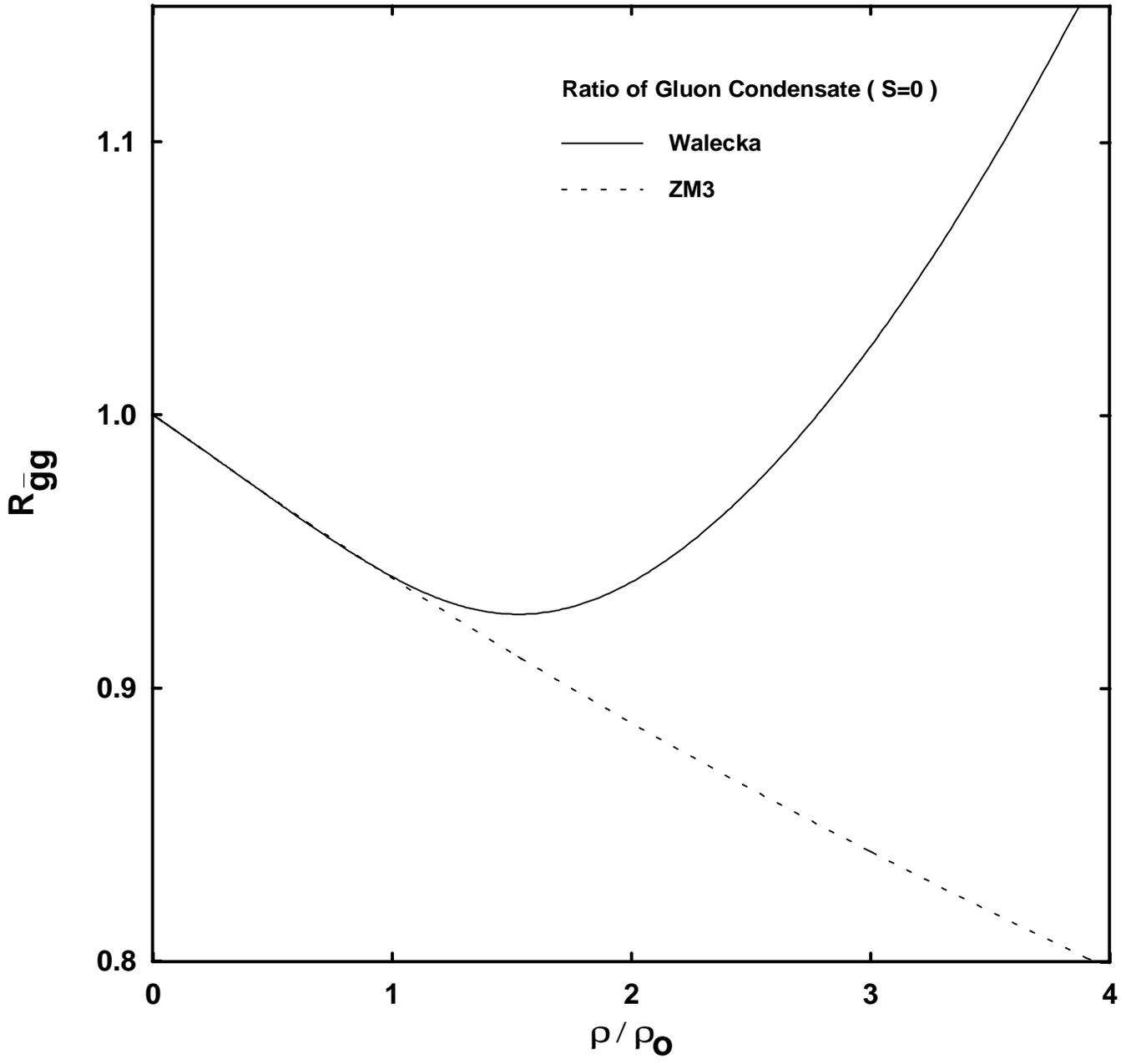
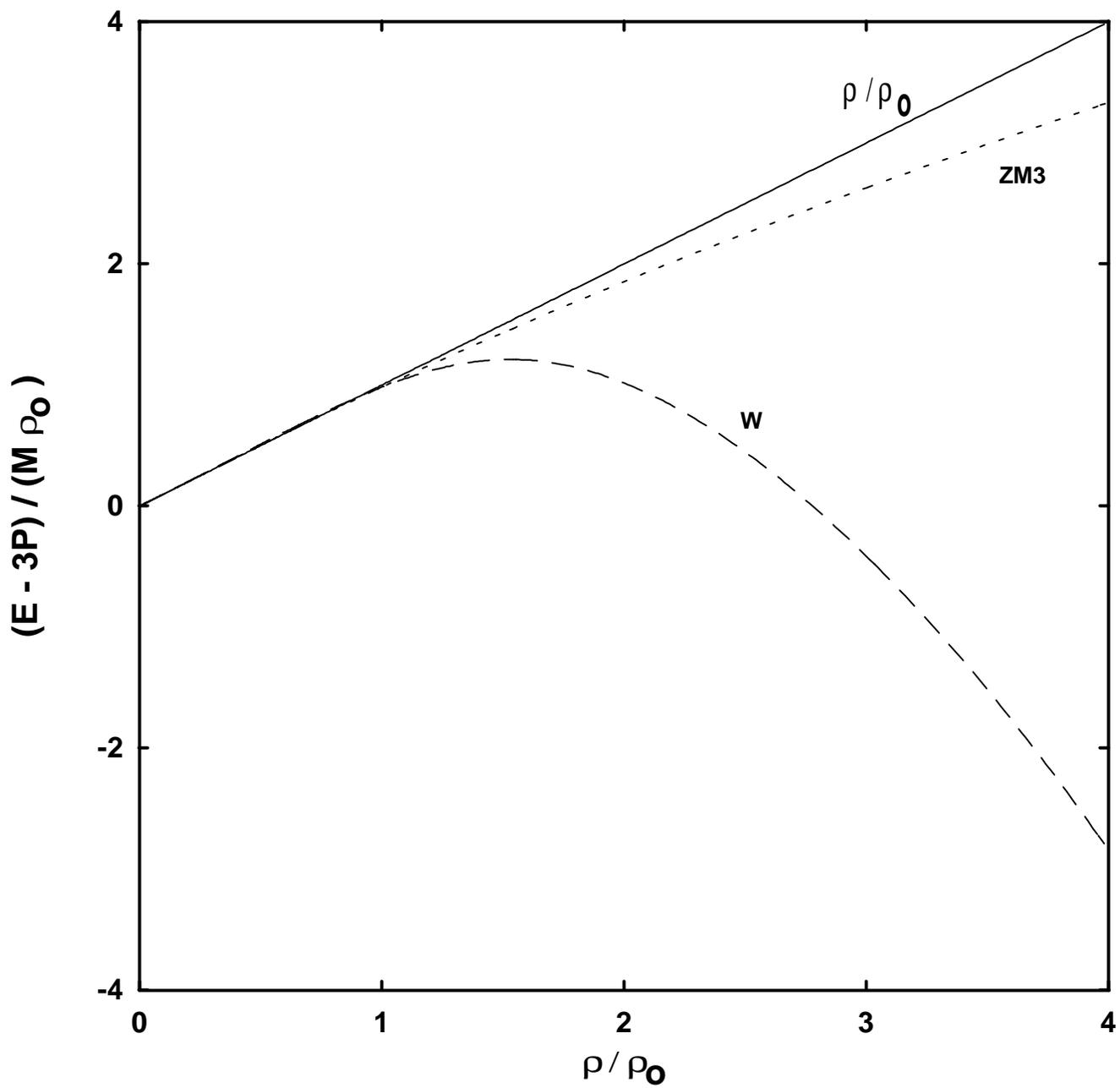
\end{document}